\documentstyle[pra,aps,preprint]{revtex}
\begin{document}
\draft
\title{TWO FERROMAGNETIC STATES \\ IN MAGNETORESISTIVE MANGANITES \\
- FIRST ORDER TRANSITION DRIVEN BY ORBITALS -}
\author{S. Maekawa, S. Ishihara and S. Okamoto}
\address{Institute for Materials Research, 
Tohoku University, 
Sendai 980-8577, Japan}
\date{July 26, 1998}
\maketitle
\begin{abstract}
A systematic study of the electronic structure in perovskite manganites 
is presented. The effective Hamiltonian is derived by taking into account 
the degeneracy of $e_g$ orbitals and strong electron correlation in Mn ions. 
The spin and orbital orderings are examined as functions of carrier concentration 
in the mean-field approximation applied to the effective Hamiltonian. We obtain 
the first order phase transition between ferromagnetic metallic and ferromagnetic 
insulating states in the lightly doped region. The transition is accompanied with 
the orbital order-disorder one which is directly observed in the anomalous X-ray 
scattering experiments. The present investigation shows a novel role of the orbital 
degree of freedom on metal-insulator transition in manganites.
\end{abstract}
\vfill
\eject

\narrowtext
\noindent
\section{Introduction}
Perovskite manganites have recently attracted much attention. They exhibit a variety 
of anomalous phenomena such as colossal magnetoresistance (CMR)\cite{r1,r2,r3,r4} and charge ordering/melting 
transitions \cite{r5,r6}, which depend crucially on carrier-doping. In this paper, we examine a 
systematic study of the electronic structure and discuss the mechanism of the properties. 
An effective Hamiltonian is derived by taking into account the degeneracy of $e_g$ orbitals, 
strong electron correlation and Hund coupling in Mn ions \cite{r7}. In the Hamiltonian, charge, spin 
and orbital degrees of freedom are included on an equal footing. The spin and orbital 
orderings are examined as functions of carrier concentration in the mean-field approximation 
applied to the effective Hamiltonian. We obtain the first order phase transition between 
ferromagnetic metallic and ferromagnetic insulating states in the lightly doped region. 
The transition is accompanied with the orbital order-disorder one. We show theoretically 
how to observe the orbital ordering by the anomalous X-ray scattering.
\par
In Sec. 2, the effective Hamiltonian is derived and examined in the mean-field approximation. 
We obtain the phase diagram as a function of carrier-doping. In Sec. 3, the anomalous X-ray 
scattering is discussed as a probe to observe the orbital ordering. The first-order phase 
transition between two ferromagnetic states is discussed in the light of the phase diagram 
in Sec. 4.  In Sec. 5, the summary and conclusion are given. The present investigation shows 
a novel role of the orbital degree of freedom on metal-insulator transition in manganites.
\section{EFFECTIVE HAMILTONIAN AND PHASE DIAGRAM}
Let us start with undoped perovskite manganites such as $\rm LaMnO_3$, where each 
$\rm Mn^{3+}$ ion has one $e_g$ and three $ t_{2g}$ electrons. 
The $ e_g$ electron occupies one of the orbitals 
$d_{3z^2-r^2}$ and $d_{x^2-y^2}$ as shown in Fig. 1, and couples with the localized $ t_{2g}$ spins ferromagnetically 
due to the Hund coupling. The Hund coupling  and the antifferomagnetic exchange 
interaction between nearest-neighbor $ t_{2g}$ spins  are given by
\begin{equation}
H_K+H_{t_{2g}}=-K \sum_i \vec S_i^{t_{2g}} \cdot \vec S_i 
+ J_s \sum _{<ij>} \vec S_i^{t_{2g}} \cdot \vec S_j^{t_{2g}} \ , 
\label{eq:eq1}
\end{equation}
where both the constants $K$ and $J_s$ are defined to be positive, 
and $\vec S^{t_{2g}}_i$ and $\vec S_i$ are the operators 
for a $t_{2g}$ spin with S=3/2 and that of $e_g$ electron with S=1/2, respectively. The 
two $e_g$ orbitals are assumed to be degenerate. The matrix elements of the electron transfer 
between $\gamma$ orbital at site $i$ and $\gamma'$ orbital at 
nearest neighbor site $j$, $t_{ij}^{\gamma \gamma'}$, is estimated by the 
second-order perturbation with respect to the electron transfer between Mn$3d$ 
and O$2p$ orbital ($t_{pd}$). $t_{pd}$ is parameterized as 
$t_{pd}=\alpha(\gamma)V_{pd \sigma}$, 
where $\alpha$ is a numerical factor and  $V_{pd\sigma}$
is an overlap integral independent of the orbitals. Then, $t_{ij}^{\gamma \gamma'}$ 
is denoted by $t_{ij}^{\gamma \gamma'}=\alpha(\gamma) \alpha(\gamma') t_0$, 
where  $t_0 (\propto V_{pd \sigma}^2)$ is treated as a parameter. 
The strong intra- and inter-orbital Coulomb 
interactions $U$ and $U'$, respectively, cause the localization of $e_g$ 
electrons. We eliminate the doubly occupied configuration in the $e_g$ 
states and obtain the leading term of the effective Hamiltonian \cite{r7}, 
\begin{equation}
H_{eff}=\widetilde H_{e_g}+H_K+H_{t_{2g}} \ , 
\label{eq:eq2}
\end{equation}
with 
\begin{equation}
\widetilde H_{e_g}=-2 \tilde J \sum_{<ij>} \Bigl( {3 \over 4} + \vec S_i \cdot \vec S_j \Bigr)
\Bigl( {1 \over 4} - \psi_i ^\dagger \hat \tau_{ij} \psi_j \Bigr) \ ,  
\label{eq:eq3}
\end{equation}
where 
\begin{equation}
\psi^\dagger_i=(T_i^z,T_i^x) \ , 
\label{eq:eq4}
\end{equation}
\begin{equation}
\hat \tau_{i i+l} ={1 \over 2}
\pmatrix { 
1+\cos n_l{2 \pi \over 3} &  \sin n_l{2 \pi \over 3} \cr
\sin n_l{2 \pi \over 3} & 1-\cos n_l{2 \pi \over 3} \cr 
} 
\label{eq:eq5}
\end{equation}
with $(n_x,n_y,n_z)=(1,2,3)$. 
Here, $\tilde J=t_0^2/(U'-J)$ with $J'(>0)$ being the intra-orbital 
exchange interaction between $e_g$ electrons.  
$\vec T_i$ is the pseudo-spin operator for the 
orbital degree of freedom at site $i$ defined as
\begin{equation}
\vec T_i={1 \over 2} \sum_{\sigma \gamma \gamma'} 
\tilde d_{i \gamma \sigma}^\dagger \vec \sigma_{\gamma \gamma'} 
\tilde d_{i \gamma' \sigma} \ , 
\label{eq:eq6}
\end{equation}
where $\vec \sigma$ is the Pauli matrix and  
$\tilde d_{i \gamma \sigma}=d_{i \gamma \sigma}
(1-n_{i \gamma -\sigma})
(1-n_{i -\gamma \sigma})
(1-n_{i -\gamma -\sigma})$
with  $d_{i \gamma \sigma}$ 
being the annihilation operator of an 
electron with spin $\sigma$ in the orbital $\gamma$ at site $i$ and 
$n_{i \gamma \sigma}=d_{i \gamma \sigma}^\dagger d_{i \gamma \sigma}$. 
The eigenstates of the operator $\vec T_i$ correspond 
to the occupied and unoccupied $e_g$ orbitals. For example, for 
the $T^z=1/2$ and $-1/2$, an 
electron occupies the $d_{3z^2-r^2}$ and $d_{x^2-y^2}$ orbitals, 
respectively. We note that Eq. (3) does not include $T^y$.
\par
In the doped manganites such as $\rm La_{1-x}Sr_xMnO_3$, eg electrons have the kinetic energy,
\begin{equation}
H_t=\sum_{<ij> \sigma \gamma \gamma'} 
(t_{ij}^{\gamma \gamma'} \tilde d_{i \gamma \sigma}^\dagger \tilde d_{i \gamma \sigma'}+ h.c.) \ . 
\label{eq:eq7} 
\end{equation}
A similar model Hamiltonian with Eq. (3) has been proposed by Khomskii and Kugel \cite{r8} 
and Castellani, Natoli and Ranninger \cite{r9}. However, 
effects of $t_{2g}$ spins were not introduced 
in their model. Roth \cite{r10}, Cyrot and Lyon-Caen \cite{r11} 
and Inagaki \cite{r12} also proposed a model, 
which coincides with ours if the matrix elements 
of the electron transfer are assumed as $t_{ij}^{\gamma \gamma'}=t_0 \delta_{\gamma \gamma'}$. 
In this case, the orbital interaction is isotropic in contrast with Eq. (3).
\par
The Hund coupling $K$ is so strong that $e_g$ and $t_{2g}$ spins at the same 
site are parallel. As seen in Eq. (3), spins prefer ferromagnetic ordering, whereas orbitals do the alternate 
ordering which is called the antiferro-type, hereafter. It is known that in the doped 
manganites, the carrier motion induces spin ferromagnetism due to the double exchange 
interaction. Therefore, we may expect a rich phase diagram as a function of the parameters ,  
and  as well as carrier concentration.
\par
We obtain the phase diagram at zero temperature as a function of hole concentration and  
in the mean field approximation \cite{r13} applied to the effective Hamiltonian. In the spin and 
orbital structures, four types of the mean field are considered; the ferro (F)-type, and 
three antiferro-types (layer(A)-type, rod(C)-type and NaCl(G)-type). We introduce the rotating 
frame in the spin and orbital spaces \cite{r14} and describe the 
states by the rotating angle $\theta_i^{(s)}$ and $\theta_i^{(t)}$, 
respectively. In the rotating frame, 
$\langle \tilde S_i^{z} \rangle 
(=\cos \theta_i^{(s)} \langle S_i^z \rangle-
  \sin \theta_i{(s)}  \langle S_i^x \rangle)$, 
and 
$\langle \tilde T_i^{z} \rangle 
(=\cos \theta_i^{(t)} \langle T_i^z \rangle
 -\sin \theta_i{(t)}  \langle T_i^x \rangle)$,
are adopted as the mean field order parameters. The kinetic energy 
term $(H_t)$ is rewritten as 
$-\sum_{<ij>} h_j^\dagger h_i \sum_\sigma z_{i \sigma}^{(s) \ast}  z_{j \sigma}^{(s)} 
\sum_{\gamma \gamma'} z_{i \gamma}^{(t) \ast} t_{ij}^{\gamma \gamma'} z_{j \gamma'}^{(t)} +h.c. $, 
where $h_i$ is a fermion operator describing the hole 
carrier and  $z_{i \sigma}^{(s)} (z_{i \gamma}^{(t)})$ 
is the element of the unitary matrix for the rotation in the spin (orbital) 
frame. We assume $ \langle h_i^\dagger h_i \rangle=x$ and represent 
$\langle z_{i \sigma}^{(s) \ast} z_{i \sigma}^{(s) } \rangle $ and 
$\langle z_{i \gamma}^{(t) \ast} t_{ij}^{\gamma \gamma'} z_{i \gamma'}^{(t) } \rangle$ 
by the rotating angle. By minimizing the energy, 
the phase diagram at $T=0$ is obtained.
\par
The results are shown in Fig. 2. The value of $J_s/t_0$ in the manganites which we are interested 
in is estimated to be $0.001 \sim 0.01$ from the Neel temperature of $\rm CaMnO_3$ \cite{r15}. 
Therefore, let us consider the case with $J_s/t_0=0.004$. At the hole concentration 
$x=0.0$, the A-type antiferromagnetic 
spin ordering is realized. With increasing $x$, the ferromagnetic spin state appears where 
the orbitals show the ordering given in Fig. 3 (a). This state is called $F_1$. With further 
increasing $x$, we find the phase separation between two ferromagnetic spin states $F_1$ and $F_2$. 
The state $F_2$ shows the orbital ordering given in Fig. 3 (b) which provides the gain of the 
kinetic energy of hole carriers. Since the state has more holes than $F_1$, the ferromagnetism 
in the state $F_2$ is caused by the double exchange interaction. On the other hand, the state 
$F_1$ has less holes and the ferromagnetism is due to the superexchange 
interaction induced by the orbital antiferro-type ordering. In Fig. 4, the total 
energy is plotted as a function of $x$. The two minima correspond to $F_1$ and $F_2$. For 
example, the compound with $x=0.2$ shows the phase separation, and $60 \%$ and $40 \%$ of 
the sample are $F_1$ with $x=0.06$ and $F_2$ with $x=0.41$, respectively. Recently, the first-order 
phase transition between two ferromagnetic spin states has been experimentally observed. 
The details will be discussed in Sec. 4. Fig. 2 also shows the phase separation between 
A-type and C-type antiferromagnetic states when  is large. Such a phase separation may 
be realized in the materials with small hopping parameter $t$. The two ferromagnetic states 
have also been shown by Maezono {\it et al}.\cite{r16}  The results are in accord with ours although the 
theoretical method is different each other. Yunoki {\it et al}.\cite{r17} have proposed the phase separation 
between two ferromagnetic states due to the Jahn-Teller coupling without electron correlation.
\section{ORBITAL ORDERING AND ANOMALOUS X-RAY SCATTERING}
As discussed in the previous section, the orbital ordering plays a crucial role 
in the magnetic and electronic properties in manganites. However, the direct observation 
of the orbitals was limited experimentally. Recently, Murakami {\it et al}. have applied 
the anomalous X-ray scattering in order to detect the orbital ordering in single 
layered manganites $\rm La_{0.5}Sr_{1.5}MnO_4$ \cite{r18}. 
They focused on a reflection at (3/4,3/4,0) point and 
obtained a resonant-like peak near the K-edge of a $\rm Mn^{3+}$ 
ion below about 200K. 
They further observed the unique polarization dependence which is attributed to 
the tensor character of the anomalous scattering factor. When all $\rm Mn^{3+}$ 
ions are equivalent, 
the reflection at (3/4,3/4,0) is forbidden. Therefore, an appearance of the intensity implies 
that two kinds of orbital are alternately aligned in the $\rm MnO_2$ plane 
(antiferro-type). 
The experimental results also imply that the dipole transition between Mn $1s$ and Mn $4p$ 
orbitals causes the scattering. The experimental method was extended to 
$\rm La_{1-x}Sr_xMnO_3$ 
with $x=0.0$ \cite{r19} and $0.12$ \cite{r20}. In this section, we study theoretically the anomalous X-ray 
scattering in relation to its role as a detector of the orbital ordering in manganites \cite{r21}.
\par
The structure factor of the X-ray scattering is expressed as a sum of the normal and 
anomalous part of the atomic scattering factor. The normal part is given by the Fourier 
transform of the charge density $\rho_i$ in the $i$-th 
atom, $f_{0i}= \langle f|\rho_i(\vec K=\vec k''-\vec k')|0 \rangle$, where $|0 \rangle(|f \rangle)$ 
is the initial (final) 
electronic state with energy $\varepsilon_0$ $(\varepsilon_f)$, and $\vec k'$ 
and $\vec k''$ are the momenta of incident and scattered 
photons, respectively. The anomalous part is derived by the interaction between electronic 
current and photon and is expressed as
\begin{equation}
\Delta f_{i \alpha \beta}=
{m \over e^2} \sum_i
\Biggl( 
 {\langle f|j_{i \alpha}(- \vec k')|l \rangle \langle l|j_{i \beta}(\vec k'')|0 \rangle \over 
 \varepsilon_0-\varepsilon_l-\omega_{k''}-i \delta }
+{\langle f|j_{i \beta}(\vec k'')|l \rangle \langle l|j_{i \alpha}(- \vec k')|0 \rangle \over 
 \varepsilon_0-\varepsilon_l+\omega_{k'}-i \delta }
\Biggr) \ , 
\label{eq:eq8}
\end{equation}
where $|l \rangle $ is the intermediate electronic states with energy $\varepsilon_l$ and $\delta$ 
is a dumping constant. 
The current operator $j_{i \alpha}(\vec k)$ describes the dipole transition between 
Mn $1s$ and $4p$ orbitals at 
site $i$ coupled with photon with polarization in the $\alpha$ direction. The contribution from 
the quadrupole transition is neglected because the inversion symmetry is preserved in 
the system which we are interested in.
\par
As mentioned above, the anomalous scattering is dominated by the Mn $1s \rightarrow 4p$ E1 
transition. In this case, how does the $3d$ orbital ordering reflect on the anisotropy 
of the anomalous scattering factor? In order to study the problem, we consider the 
electronic structure in a $\rm MnO_6$ octahedron, since the local electronic excitation 
dominates . Then, we find that the electron hybridization do not result in the 
anisotropy of the scattering factor, since the hybridization between the Mn $3d$ 
and O $2p$ orbitals and between the Mn $4p$ and O $2p$ ones are decoupled. One of the 
promising origins of the anisotropy of the scattering factor is the Coulomb 
interaction between Mn $3d$ and $4p$ electrons. The electron-electron interaction 
in the orbital ordered state breaks the cubic symmetry and thus lifts the degeneracy of Mn $4p$ orbitals.
\par
The interaction between Mn $3d$ and $4p$ electrons is represented as
\begin{equation}
V(3d_{\gamma_{\theta \pm}})=F_0+4F_2 \cos \Bigl( \theta \pm m_\gamma {2 \pi \over 3} \Bigr) \ , 
\label{eq:eq9}
\end{equation}
where $m_x=+1$, $m_y=-1$, and $m_z=0$, and 
$|3d_{\gamma_{\theta+}} \rangle=\cos(\theta/2)|3z2-r2 \rangle +\sin(\theta/2)|x2-y2 \rangle$ and 
$|3d_{\gamma_{\theta-}} \rangle$
is its counterpart. 
$F_n$ is the Slater integral between $3d$ and $4p$ electrons. The explicit formula of  
$F_n$ is given by  
$F_0=F^{(0)}$
and 
$F_2=F^{(2)}/35$
with 
$F^{(n)}=\int dr dr' r^2 r'^2 R_{3d}(r)^2 R_{4p}(r')^2 {r_{<}^n \over r_{>}^{n+1}}$, 
where 
$r_{<}$ ($r_{>}$) is the smaller (larger) one between $r$ and $r'$. 
When $d_{3z^2-r^2}$ orbital is occupied ($\theta=0$), the energy in the 
$4p_z$ orbital is higher than that of 
the $4p_x$ ($4p_y$) orbital by $6F_2$. 
As a result, $(\Delta f_i)_{xx(yy)}$ 
dominates the anomalous scattering near 
the edge in comparison with $(\Delta f_i)_{zz}$.
\par
The inter atomic Coulomb interaction between Mn $4p_\gamma$ 
electron and O $2p_{\gamma_{\theta-}}$ hole also 
provides an origin of the anisotropy of the scattering factor through the Mn 
$3d$-O $2p$ hybridization. The interaction is represented 
by $V(2p_{\gamma_{\theta-}},4p_\gamma)=-\varepsilon + {\varepsilon \rho^2 \over 5} 
\cos(\theta+m_\gamma {2 \pi \over 3})$, 
where the definition 
of $m_\gamma$ is the same as that in Eq. (9). 
$\varepsilon=Ze^2/a$ and $\rho= \langle r_{4p} \rangle/a$,  
where $Z=2$, $a$ is the 
Mn-O bond length, and $ \langle r_{4p} \rangle $ is the average radius of Mn $4p$ orbital. Although the 
above two interactions cooperate to bring about the anisotropy of the scattering 
factor, it seems likely that the magnitude of $V(2p_{\gamma_{\theta-}},4p_\gamma)$
is much reduced by the screening 
effects in comparison with $V(3d_{\gamma_{\theta+}},4p_\gamma)$.
\par
Being based on the Hamiltonian, the imaginary part of the scattering factor is 
calculated by the configuration interaction method. The calculated $(\Delta f_i)_{\alpha \alpha}$ 
near the K-edge is shown in
Fig. 5 (a), where the $d_{3z^2-r^2}$ orbital is occupied. It is noted that the edge of the lowest 
main peak corresponds to the Mn K-edge. The detailed structure away from the edge 
may become broad and be smeared out in the experiments by overlapping with other 
peaks which are not included in the calculation. In the figure, the scattering 
intensity is governed by $(\Delta f_i)_{xx}$. Owing to the core hole potential, main and satellite 
peaks are attributed to the transition from the ground state, which is mainly 
dominated by the $|3d_{\gamma_{\theta +}}^1 \rangle$ state, to the 
$|\underline{1s} 3d_{\gamma_{\theta+}}^1  
3d_{\gamma_{\theta-}}^1 4p_{x(z)}^1 \underline{2p_{\gamma _{\theta-}}}  \rangle$
and 
$|\underline{1s} 3d_{\gamma_{\theta+}}^1   4p_{x(z)}^1   \rangle $
excited states, respectively, where 
the underlines show the states occupied by holes, although the two excited states 
strongly mix with each other. Therefore, the anisotropy in the main peak is caused 
by $V(3d_{\gamma_{\theta+}}, 4p_\gamma)$ through the Mn 
$3d$-O $2p$ hybridization. As a comparison, the results in the case 
where the $d_{x^2-y^2}$ orbital is occupied are shown in Fig. 5 (b). In the figure, the anisotropy 
near the edge is entirely opposite to that in Fig. 5 (a); i.e., the scattering factor 
near the edge is governed by $(\Delta f_i'')_{zz}$, owing to the positive value of 
$V(3d_{x^2-y^2},4p_x)-V(3d_{x^2-y^2},4p_z)$.
\section{FIRST ORDER TRANSITION BETWEEN TWO FERROMAGNETIC STATES}
Recently, the first-order phase transition between ferromagnetic metallic and 
ferromagnetic insulating states has been discovered in $\rm La_{1-x}Sr_xMnO_3$ 
with $x \sim 0.12$ \cite{r20}.
In this section, we discuss the transition in the light of the phase diagram given 
in Sec. 2. Let us first review the experimental data. The electrical 
resistivity $\rho(T)$ 
shows an anomalous temperature dependence below the Curie temperature $T_C=170 K$. It 
is metallic between $T_C$ and $T_L=145 K$. However, as temperature decreases 
below $T_L$, $\rho$ 
increases rapidly and the crystal structure changes to the less distorted (pseudo cubic) 
$O^*$ phase from the distorted orthorhombic $O'$ phase \cite{r22}. A first-order transition occurs by 
applying a magnetic field between $T_C$ and $T_L$. Discontinuous jump in both 
$\rho$ and the 
magnetization curve are brought about at the critical field $H_C(T)$. The striction 
$\Delta L$ defined as $\Delta L=L(T)-L(140K)$ tends to zero above $H_C(T)$. 
With decreasing temperature, 
$H_C(T)$ decreases and at $T_C$ it goes to zero.
\par
The temperature dependence of lattice constant and magnetic Bragg reflection intensity 
were observed in the neutron scattering experiments. The results are plotted in Fig. 6. 
The ferromagnetic order parameter jumps at $T_C$ simultaneously with the $O'$ to $O^\ast$ phase 
transition upon cooling. Superlattice reflections such as $(h,k,l+1/2)$ were also observed, 
indicating the lattice modulation due to the charge ordering, which is consistent 
with the previous results \cite{r23}. 
The two ferromagnetic phases have different orbital structure. The resonance-like 
peak appears at the $(0,3,0)$ reflection in the anomalous X-ray scattering experiments, 
when the photon energy is tuned at the K-edge (6.552 KeV) in a $\rm Mn^{3+}$ ion as shown 
in Fig. 7. Besides the energy scan, the azimuthal scan around the scattering vector 
shows the angle dependence of two-fold sinusoidal symmetry, which gives rise to a 
direct evidence of the antiferro-type orbital ordering as that in the undoped system 
$\rm LaMnO_3$ \cite{r19}. 
We stress the fact that the intensity appears only below $T_L$($O^*$ phase), as 
shown in Fig. 7, where the notable lattice distortion does not exist. Therefore, 
the antiferro-type orbital ordering is not the one associated with
the Jahn-Teller type lattice distortion. Note that the spin-wave dispersion in
$\rm La_{0.88}Sr_{0.12}MnO_3$ is nearly isotropic, which is entirely different from the 
two-dimensional relation in $\rm LaMnO_3$, due to the antiferro-type orbital ordering 
of $d_{3x^2-r^2}/d_{3y^2-r^2}$, 
which is shown in Fig. 3 (c). Therefore, we anticipate that $\rm La_{0.88}Sr_{0.12}MnO_3$ 
should 
have a different orbital state, e.g. the hybridization of 
$d_{z^2-x^2(y^2-z^2)}$ and $d_{3x^2-r^2(3y^2-r^2)}$.
\par
We consider that the two ferromagnetic states observed in this compound correspond 
to $F_1$ and $F_2$ in the phase diagram in Fig. 2, and the first-order transition occurs 
between them by applying magnetic field and/or changing temperature. At high 
temperatures, the $F_2$ phase is favorable because the entropy promotes the orbital 
disordering and carrier mobility. At low temperatures, on the other hand, the $F_1$ 
phase becomes dominant and occupies the large volume fraction in the system. 
The stabilization of $F_1$ phase by an applied magnetic field is also explained 
as follows: (1) both ferromagnetic ordering and antiferro-type orbital ordering 
are confirmed to be cooperatively stabilized (2) the magnetic moment is enlarged 
by changing the dominant magnetic coupling from the double exchange interaction 
to the superexchange interaction. The first-order phase transition from ferromagnetic 
metallic to ferromagnetic insulating states is ascribed to the simultaneous transition 
of orbital order-disorder states. The phase separation occurs between these two 
ferromagnetic phases and the insulating phase dominates the system with increasing 
the magnetic field and/or decreasing temperature. The present investigations show 
a novel role of the orbital degree of freedom in the metal-insulator transition as 
a hidden parameter, unambiguously for the first time.
\section{CONCLUSION}
We have studied the magnetic and electronic structure in perovskite manganites 
taking into account the degeneracy of $e_g$ orbitals and strong electron correlation 
in Mn ions. The spin and orbital orderings were examined as functions of carrier 
concentration in the mean-field approximation applied to the effective Hamiltonian, 
which describes the low energy states. We obtained the phase separation between 
ferromagnetic metallic and ferromagnetic insulating states in the lightly doped 
region. The two ferromagnetic states have been discovered experimentally in 
$\rm La_{1-x}Sr_xMnO_3$ with $x \sim 0.12$. 
The ferromagnetic metallic state is due to the double 
exchange interaction, whereas the ferromagnetic insulating state is caused by the 
superexchange interaction coupled with the orbital degree of freedom.
The orbital degree of freedom was considered to be a hidden parameter until recently, 
since the direct observation was much limited experimentally. We have studied 
theoretically the anomalous X-ray scattering in relation to its role as a probe 
of the orbital ordering in manganites. We conclude that the orbital degree of 
freedom is not a hidden parameter but is examined together with spin and charge 
degrees of freedom of electrons in manganites.
\par
\medskip
\noindent
ACKNOWLDGEMENTS
\par
The work described in Sect. 4 was done in collaboration with Y. Endoh, K. Hirota, 
T. Fukuda, H. Kimura, H. Nojiri, and K. Kaneko at Tohoku University and 
Y. Murakami at KEK. We thank them for providing the experimental data prior 
to publication and valuable discussions. This work was supported by Priority 
Areas Grants from the Ministry of Education, Science and Culture of Japan, and 
CREST (Core Research for Evolutional Science and Technology Corporation) Japan. 
The numerical calculation was performed at the supercomputer facilities of 
Institute for Materials Research, Tohoku University and Institute of Solid State 
Physics, University of Tokyo. S. O. acknowledges the financial support of JSPS 
Research Fellowship for Young Scientists.
\vfill
\eject
\noindent
Figure captions
\par \noindent
Fig. 1. 
Schematic illustration of $d_{3z^2-r^2}$ and $d_{x^2-y^2}$ orbitals in an octahedron of O ions.
\par
\medskip
\noindent
Fig. 2:  
Theoretical phase diagram at zero temperature calculated in the mean field 
approximation as a function of the carrier concentration $(x)$ and the antiferromagnetic 
superexchange interaction $(J_s)$ between localized $t_{2g}$ spins. $J_s$ is 
normalized by $t_0$ which is 
the electron transfer intensity between neighboring $e_g$ orbitals in Mn ions, and the 
realistic value of $J_s/t_0$ is estimated to be of the order of 0.001 for the Neel temperature 
of $\rm CaMnO_3$. The two kinds of ferromagnetic phases ($F_1$ and $F_2$) with different orbital 
structures are shown and phase separation region $PS$($F_1$/$F_2$) where $F_1$ and $F_2$ 
coexist 
with different volume fractions appears between them. $A$ and $C$ imply the layer-type 
and rod-type antiferromagnetic phases, respectively, and $PS$ implies the phase 
separated region between ferromagnetic and antiferromagnetic phases, and the two 
antiferromagnetic phases.
\par
\medskip
\noindent
Fig. 3:  
Schematic illustration of the orbital orderings obtained by the theoretical 
calculation. (a) ferromagnetic insulating state ($F_1$) and (b) ferromagnetic metallic 
state ($F_2$) given in Fig. 2. The orbital structure in $\rm LaMnO_3$ is shown in (c) for comparison.
\par
\medskip
\noindent
Fig. 4:  Total energy of a function of hole concentration $x$. 
Two minima correspond to the $F_1$ and $F_2$ states.
\par
\medskip
\noindent
Fig. 5:  The imaginary part of the scattering factor $[(\Delta f''_i)_{xx} m/\pi |A_{x(z)}|^2 ]$ 
in the case where the following orbital is occupied: (a) $\theta=0 (d_{3z^2-r^2})$ and 
(b) $\theta=\pi (d_{x^2-y^2})$. The straight and broken 
lines show $(\Delta f_i'')_{xx}$ and  $(\Delta f_i'')_{zz}$, respectively. 
The origin of the energy is taken to be arbitrary. 
Here, $m$ is the mass of an electron and $A_{x(z)}$ is the coupling constant between electron 
and photon with $x(z)$ polarization of the electric field.
\par
\medskip
\noindent
Fig. 6:  Temperature dependence of (a) lattice parameter, and (b) integrated intensity 
of (2 0 0) ferromagnetic Bragg reflection measured with 14.7 meV neutrons.\cite{r20} Between 
$T_L(=145 K)$ and $T_H(=291 K)$, the crystal structure is determined to be $O'$ (orthorhombic). 
Out side of these temperatures, $O^*$ (pseudo cubic). $T_C$ is determined to be 170 K for 
$\rm La_{0.88}Sr_{0.12}MnO_3$ which is consistent with the magnetization measurement.
\par
\medskip
\noindent
Fig. 7:  (a) Energy dependence of intensities in the anomalous X-ray scattering 
experiments at the orbital ordering reflection (0 3 0) at $T=12K$ in 
$\rm La_{0.88}Sr_{0.12}MnO_3$.\cite{r20} 
The resonant energy is determined to be 6.552 KeV. The dashed curve represents 
fluorescence showing the resonant energy corresponding to the K-edge of Mn cation. 
(b) The azimuthal angle dependence of orbital ordering reflection (0 3 0). 
The solid line is two-fold squared sine curve of angular dependence. (c) Temperature 
dependence of peak intensities of orbital ordering reflection (0 3 0).
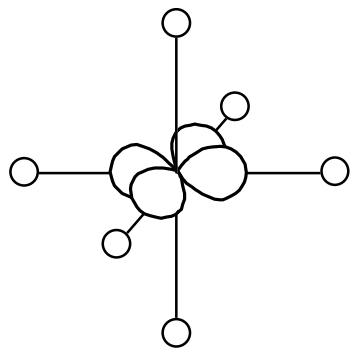
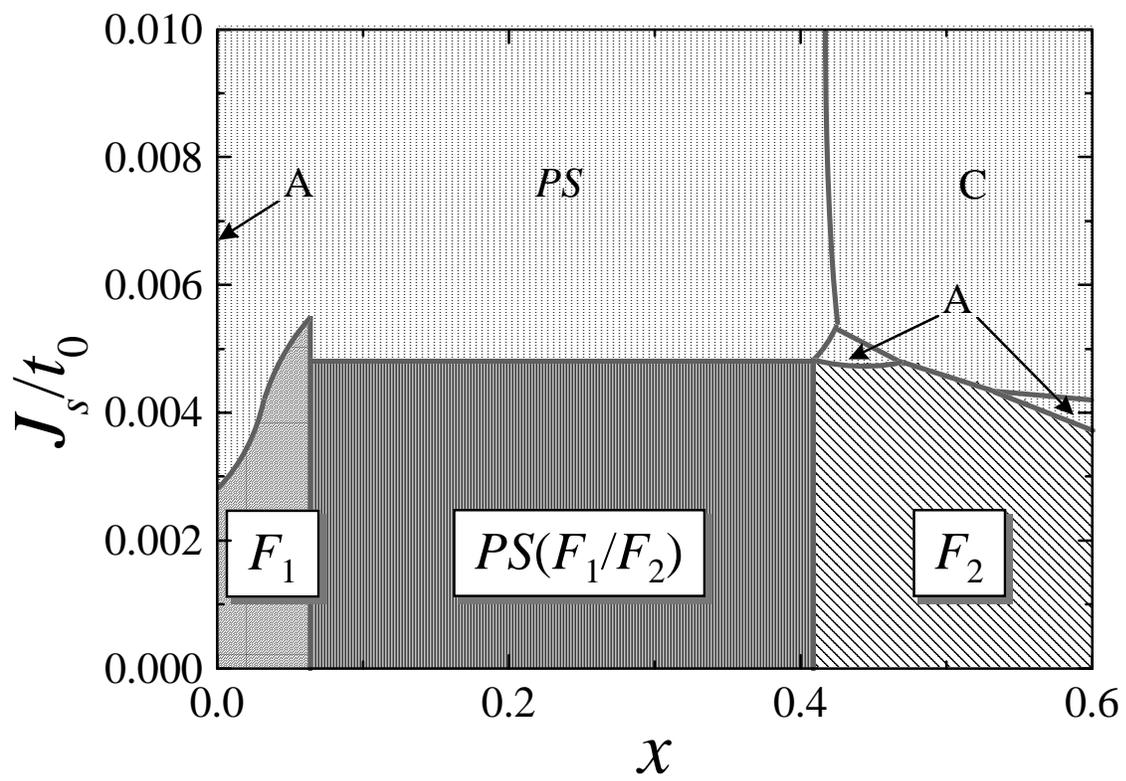
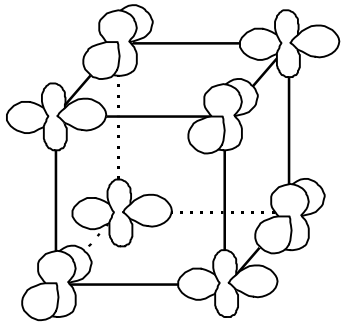
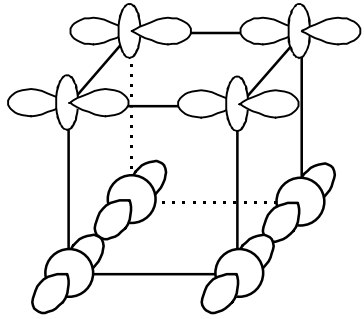
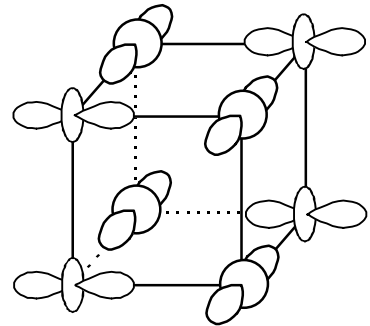
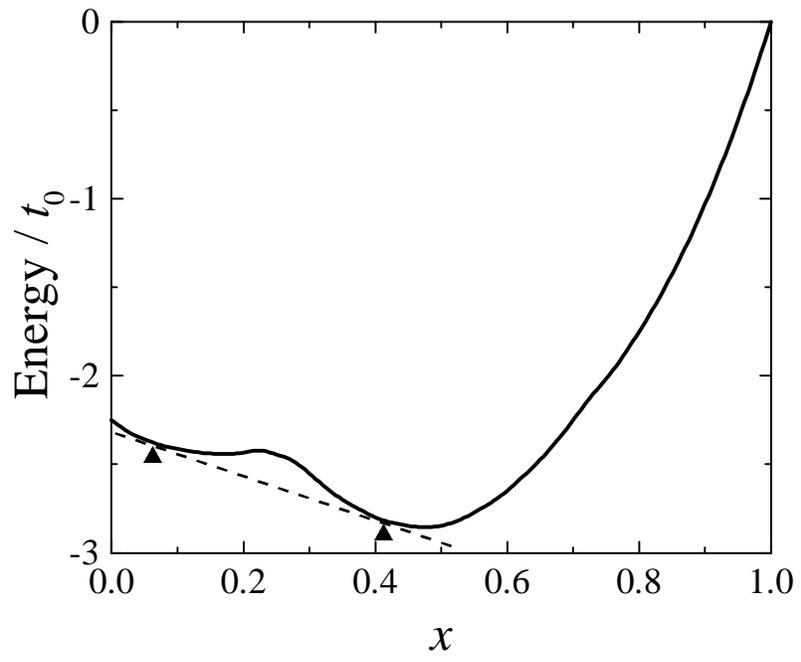
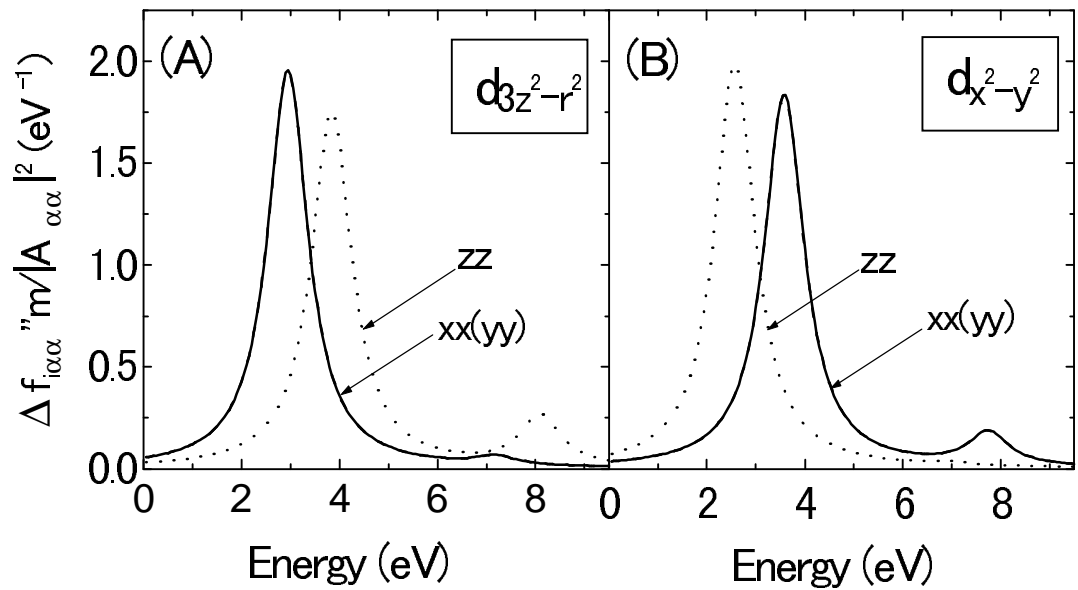
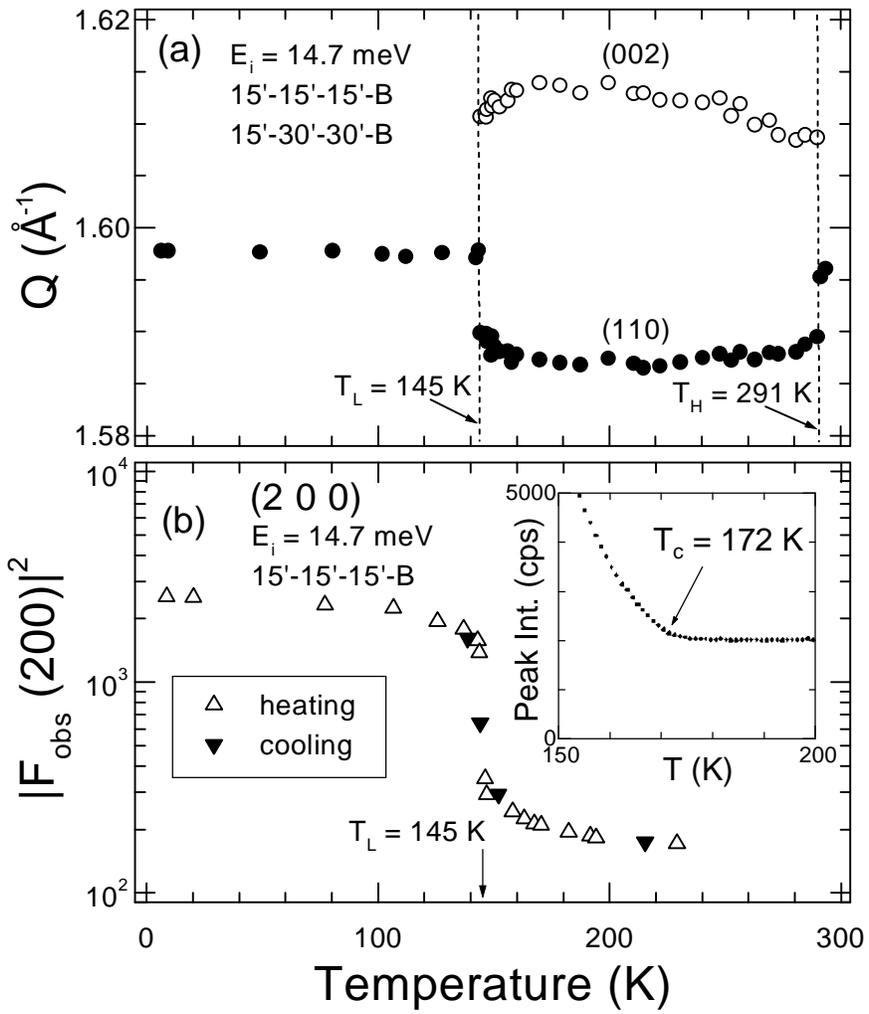
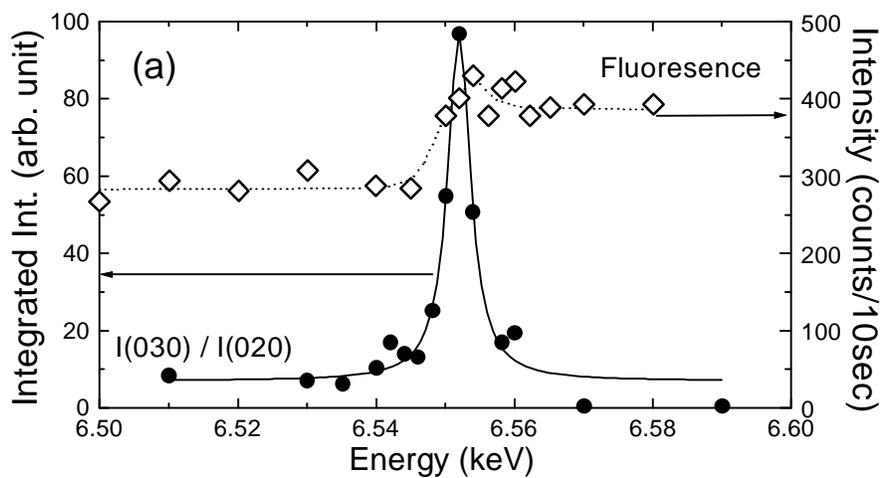
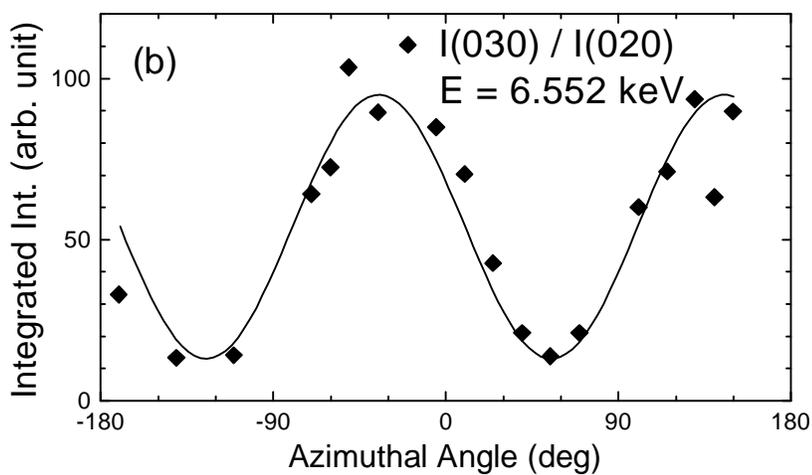
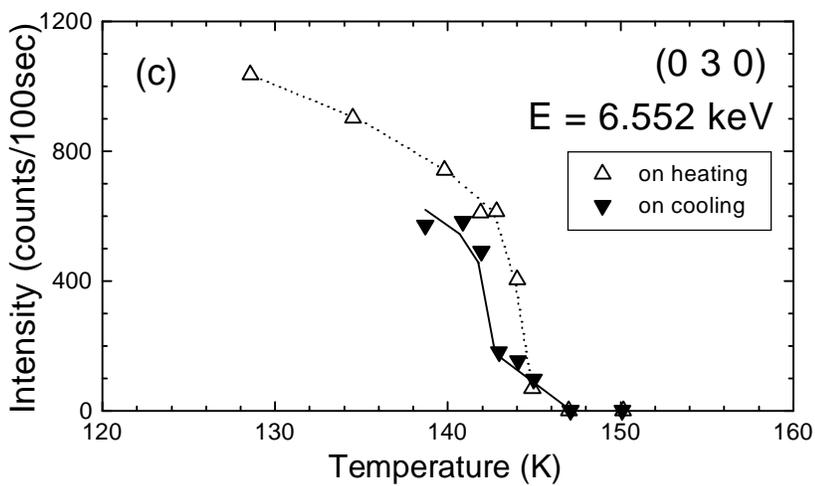
\end{document}